\documentclass[11pt,a4paper]{article}

\usepackage[margin=1in]{geometry}
\usepackage{amsmath,amssymb,amsthm}
\usepackage{hyperref}
\usepackage{cleveref}
\usepackage{enumitem}
\usepackage{braket}
\usepackage{graphicx}
\usepackage{xcolor}
\theoremstyle{definition}

\theoremstyle{remark}

\title{Quantum-Walk Inspired Spectral Metrics for Evaluating QAOA Mixers}

\author{
  Jasjyot Singh Gulati \\
  \small\textit{IIIT-Delhi, India} \\
  \small\texttt{jasjyot23257@iiitd.ac.in}
  \and
  Tharrmashastha SAPV \\
  \small\textit{EulerQ, India \& IIIT-Delhi, India} \\
  \small\texttt{tharrma@eulerq.com}
}
\date{}

\begin{document}
\maketitle

\begin{abstract}

QAOA mixers play an important role in exploring the exponentially large
solution spaces of binary optimisation problems. However, the choice of mixer
Hamiltonian is typically guided by convention and implementation or circuit
constraints, rather than by a graph-topology-aware criterion. In this work,
inspired by quantum-walk mixing theory, we propose two complementary spectral
heuristics for characterizing QAOA mixer topologies. The first is the
operator-norm-normalised minimum eigenvalue separation of the mixer Hamiltonian
$H_M$, while the second is the long-time averaged minimum eigenvalue separation
of the mixer unitary $U_M(\beta)=e^{-i\beta H_M}$. Both quantities are
invariant under positive rescaling of the mixer Hamiltonian and provide
complementary Hamiltonian- and unitary-level characterisations of its spectral
structure. We use them as heuristic indicators of the mixer's role in
amplitude redistribution during QAOA evolution.

We derive exact expressions and asymptotic bounds for the complete graph,
hypercube, cycle, and balanced 2-D torus, revealing a broad spectral separation
between the studied mixer families. We then compare this spectral
characterisation with empirical QAOA performance. For permutation-invariant
random-energy cost Hamiltonians, the observed performance ordering is broadly
consistent with the spectral predictions. In contrast, for structured 1-D and
2-D Ising QUBOs, this ordering breaks down, with the hypercube generally
matching or outperforming the complete graph despite having smaller spectral
separations. These results demonstrate both the usefulness and the limitations
of topology-level, cost-agnostic spectral heuristics for QAOA mixer selection,
and highlight the importance of mixer--cost alignment when the cost Hamiltonian
contains exploitable structure.

\end{abstract}


\section{Introduction}
\label{sec:intro}

Binary combinatorial optimisation problems are among the hardest computational problems and are commonly encountered in real-world scenarios. In the quantum setting, the Quantum Approximate Optimisation Algorithm (QAOA)~\cite{farhi2014quantum} and its generalisation, the Quantum Alternating Operator Ansatz framework~\cite{hadfield2019quantum}, are widely known variational quantum approaches aimed at solving such problems. QAOA alternates applications of a cost unitary $e^{-i\gamma H_C}$ with a mixer unitary $e^{-i\beta H_M}$ acting on a variational state $|\psi_p\rangle$. The cost unitary introduces state-dependent phases, while the subsequent mixer evolution converts these phase differences into changes in computational-basis amplitudes.
For any problem, the performance of QAOA can strongly depend on the alignment between the mixer and the structure of the cost Hamiltonian. However, the topology and spectral structure of the mixer are also important performance contributors and influence how amplitudes are redistributed across the solution space at shallow circuit depth.
Understanding the class of problems where the mixer topology largely influences the solution is an interesting line of research.

For most of the interesting problems, the mixer is not a minor implementation detail. It determines which
computational-basis states are directly connected within a single layer, and
hence how many layers are needed before amplitude can reach good solutions.
It also influences the circuit cost: the transverse-field mixer uses
only single-qubit rotations, whereas richer connectivity such as the Grover
mixer~\cite{bartschi2020grover} requires multi-controlled operations that are
costly on near-term hardware. Experimentalists therefore face a concrete
trade-off between a cheap, sparsely connected mixer and an expensive, highly
connected one, and currently this choice is made largely by convention or by
expensive trial-and-error simulation. A cheap, cost-agnostic criterion that
indicates when the extra connectivity is likely
to pay off would therefore be very useful.

Quantum walks have long been studied as a paradigm for algorithmic search and mixing. Aharonov et al.~\cite{aharonov2001quantum} established mixing-time bounds for discrete-time quantum walks with dependence on unitary eigenvalue separations. In the continuous-time setting, convergence of time-averaged quantum-walk distributions is similarly related to differences between Hamiltonian eigenvalues~\cite{Childs2025}. These observations motivate the use of Hamiltonian and unitary-level spectral separations as heuristic descriptors of mixer dynamics, without implying direct mixing-time guarantees for QAOA. Connections between quantum walks and QAOA performance have been explored \cite{matwiejew2024quantum}, as has adaptive mixer selection \cite{zhu2022adaptive}. In this work, we build on this perspective by systematically comparing canonical mixer topologies using two complementary, scale-independent spectral heuristics: a normalised Hamiltonian eigenvalue-separation metric, where the normalisation is based on the operator norm, and a time-averaged minimum unitary eigenvalue-separation metric. The latter provides a common unitary-level characterisation for mixer families with both periodic and quasiperiodic evolutions. We then empirically show that the resulting spectral characterisation agrees with empirical QAOA performance on random permutation-invariant cost Hamiltonians.

\subsection{Our Contributions}
\label{sec:intro-contributions}

In this work, we develop a topology-level spectral framework for assessing QAOA mixers
before performing costly parameter optimisation. The framework is intended
both for comparing standard mixer topologies and for screening new or
hardware-constrained mixer designs. Our results try to indicate when spectral
properties of the mixer alone are informative of QAOA performance and when
the interaction between the mixer and the structure of the cost Hamiltonian
must also be taken into account. \Cref{table:topology-comparison} summarises
the resulting spectral characterisations for the four mixer topologies
considered in this work.

\paragraph{Summary of contributions.}
\begin{itemize}

    \item \textbf{Two spectral heuristics for characterising QAOA mixers.}
    Inspired by spectral quantities that arise in quantum-walk mixing, we
    introduce two scale-independent heuristic metrics for QAOA mixers: the
    operator-norm-normalized minimum Hamiltonian eigenvalue separation
    $\widetilde{\Delta}_{H_M}$ and the long-time averaged minimum unitary
    eigenvalue separation $\overline{\delta}_{\infty}(H_M)$ of
    $U_M(\beta)=e^{-i\beta H_M}$. Together, these provide complementary
    Hamiltonian- and unitary-level characterisations of the spectral structure
    of a mixer (\cref{sec:heuristic-def}).

    \item \textbf{Analytical characterisation of canonical mixer topologies.}
    Using these metrics, we analytically characterise four canonical mixer
    topologies: the complete graph, hypercube, cycle, and balanced 2-D torus.
    We derive exact expressions or asymptotic bounds showing a clear spectral
    separation between the complete graph and hypercube on the one hand, and
    the cycle and torus on the other. In particular, the complete graph retains
    $\Theta(1)$ spectral separation, the hypercube decays only as
    $\Theta(1/\log_2 N)$, while the corresponding quantities for the cycle and
    torus decay polynomially with the size of the solution space
    (\cref{table:topology-comparison}).

    \item \textbf{Empirical agreement with the spectral predictions.}
    We then test whether the analytical spectral characterisation is reflected
    in actual QAOA performance. For permutation-invariant random cost
    Hamiltonians, where the cost function does not impose a preferred geometry
    on the computational basis, the empirical performance follows the same
    broad separation predicted by the spectral analysis: the complete-graph
    mixer performs best, followed by the hypercube, while the cycle and torus
    degrade substantially more rapidly with increasing problem size. This
    agreement provides empirical evidence that the proposed spectral quantities
    capture relevant aspects of mixer quality when the influence of
    mixer--cost alignment is absent.

    \item \textbf{Identification of the limits of cost-agnostic spectral
    characterisation.}
    Finally, we show that the same behaviour does not persist for structured
    cost Hamiltonians. On structured 1-D and 2-D Ising QUBOs, the spectral
    ordering breaks down: the hypercube generally matches or outperforms the
    complete-graph mixer despite having smaller spectral separations. This
    demonstrates that the proposed metrics are informative when the cost
    Hamiltonian is permutation-invariant, but are not sufficient on their own
    once the cost possesses exploitable structure, where mixer--cost alignment
    can become an important factor.
\end{itemize}

\subsection{Scope}
\label{sec:intro-scope}

The spectral quantities studied in this work are intended as
cost-agnostic heuristics for characterising QAOA mixer topologies, rather
than as rigorous predictors of QAOA performance. Our working hypothesis is
that, when the cost Hamiltonian does not impose a preferred structure on the
configuration space, mixers with larger spectral separations may induce more
effective amplitude redistribution and as a result exhibit better
shallow-depth QAOA performance.

We therefore use the normalised Hamiltonian eigenvalue-separation metric
$\widetilde{\Delta}_{H_M}$ and the long-time averaged minimum unitary
eigenvalue-separation metric
$\overline{\delta}_{\infty}(H_M)$ as topology-level indicators of mixer
behavior. Neither quantity is claimed to constitute a QAOA mixing-time bound
or a universal measure of mixer quality.

Their predictive value is expected to weaken when the cost Hamiltonian
introduces exploitable structure into the configuration space, since in this
regime the alignment between the mixer connectivity and the cost landscape
can become important. We test this distinction directly by comparing
permutation-invariant random-energy cost Hamiltonians with structured 1-D
and 2-D Ising QUBOs in
Section~\ref{sec:experiments-results-unstructured} and Section~\ref{sec:experiments-results-ising}.

\section{Preliminaries}
\label{sec:prelims}

\subsection{QAOA and the Quantum Alternating Operator Ansatz}
\label{sec:prelims-qaoa}

Combinatorial optimisation problems (constrained or unconstrained) over $q$ binary variables can be written as
\begin{equation}
C(x)=\sum_{a=1}^{\ell} C_a(x),
\qquad
x\in\mathcal{F}\subseteq\{0,1\}^q.
\end{equation}
Here, each $C_a(x)$ depends on a subset of the binary variables. The corresponding cost Hamiltonian is diagonal in the computational basis and is defined by
\begin{equation}
H_C
=
\sum_{x\in\mathcal{F}}
C(x)\ket{x}\bra{x}
=
\sum_{a=1}^{\ell} H_{C_a},
\qquad
H_{C_a}
=
\sum_{x\in\mathcal{F}}
C_a(x)\ket{x}\bra{x}.
\end{equation}
Since the component cost Hamiltonians $H_{C_a}$ are diagonal and therefore commute, the corresponding cost unitary can be written as
\begin{equation}
U_C(\gamma)
=
e^{-i\gamma H_C}
=
\prod_{a=1}^{\ell}
e^{-i\gamma H_{C_a}}.
\end{equation}

In the original QAOA formulation, the feasible space is the complete computational basis, $\mathcal{F}=\{0,1\}^q$, and a transverse-field mixer such as $H_M=\sum_{j=1}^{q}X_j$ is commonly used. The Quantum Alternating Operator Ansatz generalises this construction by allowing problem-dependent mixer Hamiltonians and restricting the evolution to a feasible subspace $\mathcal{F}\subseteq\{0,1\}^q$~\cite{hadfield2019quantum}. This allows for the design of mixers with connectivity tailored to the structure of the feasible solution space. In this work, we adopt this graph-based mixer perspective and represent $H_M$ through the adjacency matrix of a graph whose vertices correspond to the feasible computational-basis states.

For the unweighted, undirected mixer graphs considered here~\cite{fuchs2022constraint},
\begin{equation}
H_M
=
\sum_{x,y\in\mathcal{F}}
(H_M)_{xy}\ket{x}\bra{y},
\qquad
U_M(\beta)
=
e^{-i\beta H_M}.
\end{equation}

Starting from the uniform superposition over the feasible subspace,
\begin{equation}
\ket{\psi_0}
=
\frac{1}{\sqrt{|\mathcal{F}|}}
\sum_{y\in\mathcal{F}}\ket{y},
\end{equation}
a depth-$p$ QAOA state is prepared by alternating applications of the cost
and mixer unitaries:
\begin{equation}
\ket{\psi_p}
=
\prod_{k=1}^{p}
U_M(\beta_k)U_C(\gamma_k)\ket{\psi_0}.
\end{equation}

The variational parameters
$\boldsymbol{\gamma}=(\gamma_1,\ldots,\gamma_p)$ and
$\boldsymbol{\beta}=(\beta_1,\ldots,\beta_p)$
are then optimised to minimise the expected cost,
\begin{equation}
M_p =
\min_{\boldsymbol{\gamma},\boldsymbol{\beta}}
\langle \psi_p | H_C | \psi_p \rangle.
\end{equation}

This graph-based view of the mixer naturally motivates a connection with quantum-walk mixing, which we review next.

\subsection{Quantum Walk Mixing}
\label{sec:spectral-mixing}
\label{sec:metric}

When the mixer Hamiltonian $H_M$ is the adjacency matrix of a graph whose vertices
represent computational-basis states, the mixer evolution
$e^{-i\tau H_M}$ corresponds to a continuous-time quantum walk on that graph.
The spectral properties of the underlying Hamiltonian therefore govern
important aspects of this evolution and motivate asking whether quantities
used to characterise quantum-walk mixing can also provide useful information
about QAOA mixers.

To motivate the spectral quantities considered in this work, we briefly
recall how mixing is characterised in quantum walks. Unlike classical random
walks, unitary quantum-walk evolution does not generally converge to a
stationary probability distribution because of persistent interference.
Mixing is therefore typically described in terms of the corresponding
time-averaged probability distribution.

For a discrete-time quantum walk described by a unitary operator $U$, let
\[
P_t(v)=|\langle v|U^t|\psi_0\rangle|^2,
\qquad
\overline{P}_T(v)=\frac{1}{T}\sum_{t=0}^{T-1}P_t(v),
\]
and let
\[
\Pi(v)=\lim_{T\to\infty}\overline{P}_T(v).
\]
A standard time-averaged $\epsilon$-mixing time is then defined as the time
beyond which the averaged distribution remains within $\epsilon$ of its
limiting distribution:
\[
\tau_\epsilon
=
\min\left\{
T:
\|\overline{P}_{T'}-\Pi\|_1\leq\epsilon
\ \text{for all }T'\geq T
\right\}.
\]

The connection between this convergence and the spectrum of the walk operator
can be seen from the spectral decomposition
\[
U=\sum_j e^{i\theta_j}|\phi_j\rangle\langle\phi_j|.
\]
During the evolution, interference between different eigenspaces is governed
by differences between their eigenphases. For two eigenvalues
$e^{i\theta_j}$ and $e^{i\theta_k}$, their Euclidean separation on the unit
circle is
\[
\left|e^{i\theta_j}-e^{i\theta_k}\right|
=
2\left|
\sin\left(\frac{\theta_j-\theta_k}{2}\right)
\right|.
\]

These eigenvalue separations are known to affect the mixing behaviour of
discrete-time quantum walks~\cite{aharonov2001quantum,kargin2010bounds}.
In general, smaller separations can lead to slower convergence of the
time-averaged distribution. The precise mixing behaviour can also depend on
the eigenvectors and on the initial state.

For continuous-time quantum walks, the evolution is generated directly by a
Hamiltonian $H$:
\[
|\psi(\tau)\rangle=e^{-iH\tau}|\psi_0\rangle.
\]
Writing
\[
H=\sum_j\lambda_j|\phi_j\rangle\langle\phi_j|,
\]
the interference between different eigenspaces contains terms of the form
\[
e^{-i(\lambda_j-\lambda_k)\tau}.
\]
The evolution therefore depends naturally on the differences between
Hamiltonian eigenvalues. Smaller eigenvalue separations correspond to more
slowly varying interference terms and can lead to slower convergence of the
time-averaged distribution~\cite{Childs2025}.
This provides the continuous-time counterpart of the unitary eigenvalue
separations discussed above.

These two quantum-walk viewpoints motivate the two spectral quantities studied in this work. The Hamiltonian spectrum of $H_M$ provides a continuous-time characterisation of the mixer generator, while the spectrum of
\[
U_M(\beta)=e^{-i\beta H_M}
\]
describes how those Hamiltonian eigenvalues are mapped onto the unit circle during finite-angle mixer evolution. We therefore use Hamiltonian eigenvalue separations and unitary eigenvalue separations as complementary spectral heuristics for mixer topology.

These quantum-walk results motivate the spectral quantities considered in
this work, but they do not directly provide mixing-time bounds for QAOA.
In QAOA, the cost unitary first introduces state-dependent phases, which are
then converted by the mixer into changes in computational-basis amplitudes.
We therefore use the quantum-walk results only as motivation for spectral
descriptors of the mixer and evaluate their relevance to QAOA performance
empirically.

%%%%%%%%%%%%%%%%%%%%%%%%%%%%%%%%%%%%%%%%%%%%%%%%%%%
%%%%% THE QUANTUM WALK BASED METRICS SECTION %%%%%%
%%%%%%%%%%%%%%%%%%%%%%%%%%%%%%%%%%%%%%%%%%%%%%%%%%%

\section{Quantum-Walk-Based Spectral Metrics}
\label{sec:heuristic-def}

We characterise the mixer from two complementary spectral perspectives: the spectrum of the mixer Hamiltonian and the spectrum of the mixer unitary. Let
\begin{equation}
H_M=\sum_j\lambda_j\ket{\phi_j}\bra{\phi_j},
\qquad
U_M(\beta)=e^{-i\beta H_M}
=\sum_j e^{-i\beta\lambda_j}\ket{\phi_j}\bra{\phi_j}.
\end{equation}

\paragraph{Hamiltonian characterization.}
Since the mixer angle is variational, the overall scale of a mixer Hamiltonian is arbitrary. In particular, for any $a>0$,
\[
e^{-i\beta(aH_M)}=e^{-i(a\beta)H_M},
\]
so $H_M$ and $aH_M$ generate the same family of mixer unitaries up to a reparameterization of $\beta$. Therefore, any spectral metric intended to characterise the mixer topology should not depend on this arbitrary scale.

To ensure this, we normalise the mixer Hamiltonian by its operator norm,
\[
\widetilde{H}_M=\frac{H_M}{\|H_M\|_2}.
\]
If $\lambda_j$ are the eigenvalues of $H_M$, then the eigenvalues of $\widetilde{H}_M$ are
\[
\widetilde{\lambda}_j=\frac{\lambda_j}{\|H_M\|_2}.
\]
We define the normalised minimum nonzero Hamiltonian eigenvalue separation as
\[
\widetilde{\Delta}_{H_M}
=
\min_{\widetilde{\lambda}_i\neq\widetilde{\lambda}_j}
|\widetilde{\lambda}_i-\widetilde{\lambda}_j|
=
\frac{\Delta_{H_M}}{\|H_M\|_2}.
\]
By construction,
\[
\widetilde{\Delta}_{aH_M}
=
\widetilde{\Delta}_{H_M},
\]
so this quantity characterises spectral separation independently of the overall Hamiltonian scale.

\paragraph{Unitary characterisation.}
At each QAOA layer, the mixer is applied through a finite-angle evolution
\[
U_M(\beta_j)=e^{-i\beta_j H_M},
\]
interleaved with cost evolutions
\[
U_C(\gamma_j)=e^{-i\gamma_j H_C}.
\]
While $\widetilde{\Delta}_{H_M}$ characterises the scale-normalised spectrum of the mixer Hamiltonian, QAOA acts directly through the eigenvalues of the corresponding unitary. This motivates a complementary characterisation of the spectrum after it is mapped onto the unit circle.

For this unitary-level characterisation, we retain the canonical adjacency mixer Hamiltonian $H_M$ rather than its normalised form. For a fixed mixer angle $\beta$, we define the minimum unitary eigenvalue separation as
\[
\delta_{H_M}(\beta)
=
\min_{\lambda_i\neq\lambda_j}
\left|
e^{-i\beta\lambda_i}
-
e^{-i\beta\lambda_j}
\right|,
\]
or equivalently,
\[
\delta_{H_M}(\beta)
=
2\min_{\lambda_i\neq\lambda_j}
\left|
\sin\left(
\frac{\beta(\lambda_i-\lambda_j)}{2}
\right)
\right|.
\]

While $\widetilde{\Delta}_{H_M}$ measures eigenvalue separation directly in
the mixer Hamiltonian, $\delta_{H_M}(\beta)$ measures the separation between
the corresponding eigenvalues of the mixer unitary at angle $\beta$. This
captures effects such as phase wrapping and eigenvalue collisions that are
not visible from the Hamiltonian spectrum alone.

Because $\beta$ is a variational parameter and there is no distinguished mixer angle at which a topology should be evaluated, a metric based on a single value of $\beta$ would be arbitrary. Similarly, averaging over a fixed finite interval would introduce an additional externally chosen scale. We therefore define the intrinsic, $\beta$-independent quantity
\[
\boxed{
\overline{\delta}_{\infty}(H_M)
=
\lim_{B\to\infty}
\frac{1}{B}
\int_0^B
\delta_{H_M}(\beta)\,d\beta
}.
\]
For the finite spectra considered in this work, $\delta_{H_M}(\beta)$ is a continuous almost-periodic function of $\beta$, so the above long-time mean exists. The definition therefore treats periodic and quasiperiodic mixer evolutions within the same framework.

If the mixer unitary possesses a finite projective period $T_M$, i.e.,
\[
U_M(\beta+T_M)=e^{i\phi}U_M(\beta)
\]
for a global phase $\phi$ independent of $\beta$, then $\delta_{H_M}(\beta)$ is periodic with period $T_M$, and therefore
\[
\overline{\delta}_{\infty}(H_M)
=
\frac{1}{T_M}
\int_0^{T_M}
\delta_{H_M}(\beta)\,d\beta.
\]
Thus, for periodic mixers the long-time average reduces to the average over one complete projective period, while for mixers without a finite common projective period it provides the corresponding asymptotic time average without introducing an arbitrary finite averaging window.

The long-time averaged unitary metric is also invariant under a positive rescaling of the mixer Hamiltonian. For $a>0$,
\[
\delta_{aH_M}(\beta)=\delta_{H_M}(a\beta),
\]
and hence
\[
\begin{aligned}
\overline{\delta}_{\infty}(aH_M)
&=
\lim_{B\to\infty}
\frac{1}{B}
\int_0^B
\delta_{H_M}(a\beta)\,d\beta \\
&=
\lim_{B\to\infty}
\frac{1}{aB}
\int_0^{aB}
\delta_{H_M}(u)\,du \\
&=
\overline{\delta}_{\infty}(H_M).
\end{aligned}
\]
Thus, both $\widetilde{\Delta}_{H_M}$ and $\overline{\delta}_{\infty}(H_M)$ are independent of the arbitrary overall scale of the mixer Hamiltonian.

Having established the two spectral metrics, we next derive a simple general
bound on the unitary metric that depends only on the number of distinct
eigenvalues of the mixer Hamiltonian.

\paragraph{A general bound from the number of distinct eigenvalues.}
Let $m(H_M)$ denote the number of distinct eigenvalues of $H_M$. Assume $m(H_M)\geq 2$. For any fixed $\beta$, the corresponding unitary eigenvalues lie on the unit circle. If two distinct Hamiltonian eigenvalues map to the same unitary eigenvalue, then $\delta_{H_M}(\beta)=0$ and the following bound is immediate. Otherwise, the $m(H_M)$ unitary eigenvalues form $m(H_M)$ distinct points on the unit circle. The angular gaps between consecutive points sum to $2\pi$, so at least one such gap is at most $2\pi/m(H_M)$. Since a chord subtending an angle $\theta$ on the unit circle has length $2\sin(\theta/2)$,
\[
\delta_{H_M}(\beta)
\leq
2\sin\left(
\frac{\pi}{m(H_M)}
\right).
\]
Consequently,
\[
\boxed{
\overline{\delta}_{\infty}(H_M)
\leq
2\sin\left(
\frac{\pi}{m(H_M)}
\right)
}.
\]
This bound will be useful below for the hypercube, cycle, and torus mixer families.

\paragraph{Interpretation and scope.}
We use $\widetilde{\Delta}_{H_M}$ and
$\overline{\delta}_{\infty}(H_M)$ as topology-level spectral heuristics for
QAOA mixers, rather than as rigorous QAOA mixing-time bounds. Our working
hypothesis is that, when the cost Hamiltonian does not impose a preferred
geometry on the configuration space, mixers with larger spectral separations
may exhibit better shallow-depth QAOA performance. We test this relationship
empirically in the following sections.

The proposed metrics depend only on the eigenvalues of the mixer Hamiltonian
and therefore do not capture mixer eigenvectors, detailed basis-state
connectivity, or the structure of the cost Hamiltonian. Their predictive value
is consequently expected to weaken when mixer--cost alignment becomes
important, particularly for structured cost Hamiltonians. This limitation is
discussed further in \cref{sec:intro-scope}.

\section{Analytic Results}

We now evaluate the two spectral metrics introduced in Section~\ref{sec:heuristic-def} for four canonical mixer topologies: the complete graph, hypercube, cycle, and balanced 2-D torus. Where possible, we derive exact expressions or matching asymptotic bounds. For the cycle and torus, some quantities admit only upper bounds, and we state these limitations explicitly rather than inferring an asymptotic ordering that has not been established.

\subsection{Complete Graph \texorpdfstring{$K_N$}{KN}}
\label{sec:analytic-complete}

For the complete graph $K_N$, the two metrics are straightforward to derive. The distinct eigenvalues of the adjacency Hamiltonian are $\lambda\in\{N-1,-1\}$. Since $\|H_M\|_2=N-1$, the normalized Hamiltonian has distinct eigenvalues $1$ and $-\frac{1}{N-1}$. Hence the normalised minimum Hamiltonian eigenvalue separation is
\[
\widetilde{\Delta}_{K_N}
=
1+\frac{1}{N-1}
=
\frac{N}{N-1}
=
\Theta(1).
\]

For the unitary characterisation, we retain the canonical adjacency Hamiltonian $H_M$. Since its two distinct eigenvalues differ by $N$, the unitary eigenvalue separation is
\[
\delta_{K_N}(\beta)
=
2\left|
\sin\left(\frac{N\beta}{2}\right)
\right|.
\]
The corresponding mixer unitary is projectively periodic with fundamental period $T_{K_N}=2\pi/N$. By the periodic reduction established in Section~\ref{sec:heuristic-def}, the long-time average is therefore equal to the average over one complete projective period:
\[
\begin{aligned}
\overline{\delta}_{\infty}(K_N)
&=
\frac{1}{T_{K_N}}
\int_0^{T_{K_N}}
\delta_{K_N}(\beta)\,d\beta \\
&=
\frac{N}{2\pi}
\int_0^{2\pi/N}
2\left|
\sin\left(\frac{N\beta}{2}\right)
\right|\,d\beta \\
&=
\frac{4}{\pi}
=
\Theta(1).
\end{aligned}
\]

\subsection{Hypercube \texorpdfstring{$Q_q$}{Qq}}
\label{sec:analytic-hypercube}

Let $Q_q$ be the $q$-dimensional hypercube, with $N=2^q$. Its distinct eigenvalues are $\lambda_k=q-2k$ for $k=0,\ldots,q$. Since $\|H_M\|_2=q$, the normalized eigenvalues are $\widetilde{\lambda}_k=1-\frac{2k}{q}$, and hence the normalized minimum Hamiltonian eigenvalue separation is
\[
\widetilde{\Delta}_{Q_q}
=
\frac{2}{q}
=
\Theta\left(\frac{1}{q}\right)
=
\Theta\left(\frac{1}{\log_2 N}\right).
\]

For the unitary characterisation, any difference between two distinct eigenvalues can be written as $2t$, where $t=|\ell-k|\in\{1,\ldots,q\}$. Therefore,
\[
\delta_{Q_q}(\beta)
=
2\min_{t=1,\ldots,q}|\sin(t\beta)|.
\]
Since all eigenvalue differences are even integers, the mixer unitary is projectively periodic with fundamental period $T_{Q_q}=\pi$. Hence its long-time average is
\[
\overline{\delta}_{\infty}(Q_q)
=
\frac{1}{\pi}
\int_0^\pi
\delta_{Q_q}(\beta)\,d\beta.
\]

The hypercube has
\[
m(Q_q)=q+1
\]
distinct Hamiltonian eigenvalues. Applying the general spectral-cardinality bound from Section~\ref{sec:heuristic-def},
\[
\delta_{Q_q}(\beta)
\leq
2\sin\left(\frac{\pi}{q+1}\right),
\]
and therefore
\[
\overline{\delta}_{\infty}(Q_q)
\leq
2\sin\left(\frac{\pi}{q+1}\right)
=
O\left(\frac{1}{q}\right).
\]

To obtain a matching lower bound, let
$X=\min_{t=1,\ldots,q}|\sin(t\beta)|$ for
$\beta\sim U[0,\pi]$, so that
$\delta_{Q_q}(\beta)=2X$. Since $0\leq X\leq1$, the tail-integral representation of the expectation gives
\[
\overline{\delta}_{\infty}(Q_q)
=
2\mathbb{E}[X]
=
2\int_0^1 \Pr(X>s)\,ds.
\]

The event $X\leq s$ occurs whenever at least one of the $q$ terms $|\sin(t\beta)|$ is at most $s$. Hence, by the union bound,
\[
\Pr(X\leq s)
\leq
\sum_{t=1}^{q}
\Pr\bigl(|\sin(t\beta)|\leq s\bigr).
\]
Because $\beta$ is uniformly distributed over $[0,\pi]$ and $|\sin x|$ has period $\pi$, the function $|\sin(t\beta)|$ traverses exactly $t$ complete periods as $\beta$ ranges over $[0,\pi]$. Consequently, for every integer $t\geq1$, the fraction of angles for which $|\sin(t\beta)|\leq s$ is the same as for $|\sin\beta|$. Within one period, this condition holds on intervals of total length $2\arcsin(s)$, so
\[
\Pr\bigl(|\sin(t\beta)|\leq s\bigr)
=
\frac{2}{\pi}\arcsin(s),
\]
and therefore
\[
\Pr(X\leq s)
\leq
\frac{2q}{\pi}\arcsin(s).
\]

Since
\[
\Pr(X>s)=1-\Pr(X\leq s),
\]
the union bound yields a nontrivial lower bound on $\Pr(X>s)$ whenever
$\frac{2q}{\pi}\arcsin(s)\leq1$, i.e.,
$s\leq\sin(\frac{\pi}{2q})$. Restricting the tail integral to this range therefore gives
\[
\begin{aligned}
\overline{\delta}_{\infty}(Q_q)
&\geq
2\int_0^{\sin(\pi/2q)}
\left[
1-\frac{2q}{\pi}\arcsin(s)
\right]ds \\
&=
\frac{4q}{\pi}
\left(
1-\cos\frac{\pi}{2q}
\right).
\end{aligned}
\]

Using $1-\cos x\geq \frac{2}{\pi^2}x^2$ for $x\in[0,\pi]$, we obtain
\[
\overline{\delta}_{\infty}(Q_q)
\geq
\frac{2}{\pi q}
=
\Omega\left(\frac{1}{q}\right)
=
\Omega\left(\frac{1}{\log_2 N}\right).
\]

Combining the upper and lower bounds,
\[
\boxed{
\overline{\delta}_{\infty}(Q_q)
=
\Theta\left(\frac{1}{q}\right)
=
\Theta\left(\frac{1}{\log_2 N}\right)
}.
\]

\subsection{Cycle Graph \texorpdfstring{$C_N$}{CN}}
\label{sec:analytic-cycle}

For the cycle graph $C_N$, the eigenvalues of the adjacency Hamiltonian are $\lambda_k=2\cos(2\pi k/N)$. Since $\lambda_k=\lambda_{N-k}$, the distinct eigenvalues may be indexed by
$k\in\{0,\ldots,\lfloor N/2\rfloor\}$.
Over this range, $2\pi k/N\in[0,\pi]$, on which $\cos\theta$ is strictly decreasing, so these indexed values are indeed distinct. The separation between adjacent distinct eigenvalues is
\[
|\lambda_k-\lambda_{k+1}|
=
4\sin\left(\frac{\pi}{N}\right)
\sin\left(\frac{\pi(2k+1)}{N}\right).
\]
The minimum is attained, in particular, at $k=0$, giving
\[
\Delta_{C_N}
=
|\lambda_0-\lambda_1|
=
4\sin^2\left(\frac{\pi}{N}\right)
=
\Theta\left(\frac{1}{N^2}\right).
\]
Since $C_N$ is $2$-regular, $\|H_M\|_2=2$, and hence the normalized Hamiltonian eigenvalue separation is
\[
\widetilde{\Delta}_{C_N}
=
\frac{\Delta_{C_N}}{2}
=
2\sin^2\left(\frac{\pi}{N}\right)
=
\Theta\left(\frac{1}{N^2}\right).
\]

For the unitary characterisation, the cycle does not in general possess a finite common projective period. We therefore use the long-time average $\overline{\delta}_{\infty}(C_N)$ introduced in Section~\ref{sec:heuristic-def}.

The number of distinct cycle eigenvalues is
\[
m(C_N)
=
\left\lfloor\frac{N}{2}\right\rfloor+1.
\]
In the setting considered here, $N=2^q$ is even, and hence
\[
m(C_N)=\frac{N}{2}+1.
\]
Applying the general spectral-cardinality bound,
\[
\delta_{C_N}(\beta)
\leq
2\sin\left(
\frac{\pi}{N/2+1}
\right)
=
2\sin\left(
\frac{2\pi}{N+2}
\right)
\]
for every $\beta$. Consequently,
\[
\boxed{
\overline{\delta}_{\infty}(C_N)
\leq
2\sin\left(
\frac{2\pi}{N+2}
\right)
=
O\left(\frac{1}{N}\right)
}.
\]
This is an upper bound on the long-time averaged unitary separation; a corresponding asymptotic lower bound is not established here.

\subsection{2-D Torus \texorpdfstring{$T_{r,c}$}{Trc}}
\label{sec:analytic-torus}

For the experiments considered in this work, $N=2^q$. We choose the torus dimensions $r=2^{\lfloor q/2\rfloor}$ and $c=2^{\lceil q/2\rceil}$, so that $rc=N$ and the torus is as balanced as possible. Consequently, $r,c=\Theta(\sqrt{N})$, and in particular $1/r^2=\Theta(1/N)$.

The eigenvalues of the adjacency Hamiltonian of $T_{r,c}$ are
\[
\lambda_{a,b}
=
2\cos\left(\frac{2\pi a}{r}\right)
+
2\cos\left(\frac{2\pi b}{c}\right),
\]
where $a\in\{0,\ldots,r-1\}$ and $b\in\{0,\ldots,c-1\}$.

The true minimum spectral spacing is obtained by minimising over all distinct mode pairs $(a,b)\neq(a',b')$. Therefore, evaluating any specific pair provides an upper bound. Since $T_{r,c}$ is $4$-regular, $\|H_M\|_2=4$. Choosing the modes $(0,0)$ and $(1,0)$ gives
\[
\widetilde{\Delta}_{T_{r,c}}
\leq
\frac{1}{4}
|\lambda_{0,0}-\lambda_{1,0}|
=
\sin^2\left(\frac{\pi}{r}\right)
=
O\left(\frac{1}{N}\right).
\]

We emphasise that this is only an upper bound on the normalised minimum distinct-eigenvalue separation. A matching lower bound would require controlling differences of the form
\[
\lambda_{a,b}-\lambda_{a',b'}
=
2\left(
\cos\frac{2\pi a}{r}
-
\cos\frac{2\pi a'}{r}
\right)
+
2\left(
\cos\frac{2\pi b}{c}
-
\cos\frac{2\pi b'}{c}
\right),
\]
where the two contributions may partially cancel. Consequently, the true minimum spacing may be smaller than the candidate spacing above, and determining its exact asymptotic scaling is left open.

For the unitary characterisation, the torus also does not in general possess a finite common projective period. We therefore use the same long-time average $\overline{\delta}_{\infty}(T_{r,c})$ as for all other mixer families.

Let
\[
\Lambda_r
=
\left\{
2\cos\left(\frac{2\pi a}{r}\right)
:
a=0,\ldots,r-1
\right\}_{\mathrm{distinct}}
\]
and
\[
\Lambda_c
=
\left\{
2\cos\left(\frac{2\pi b}{c}\right)
:
b=0,\ldots,c-1
\right\}_{\mathrm{distinct}}.
\]

For a cycle $C_r$, the eigenvalues indexed by $a$ and $r-a$ are equal, since
\[
\cos\left(\frac{2\pi a}{r}\right)
=
\cos\left(\frac{2\pi(r-a)}{r}\right).
\]
Thus it is sufficient to consider $a=0,\ldots,\lfloor r/2\rfloor$. These values are all distinct because $2\pi a/r\in[0,\pi]$ over this range and $\cos\theta$ is strictly decreasing on $[0,\pi]$. Hence
\[
|\Lambda_r|
=
\left\lfloor\frac{r}{2}\right\rfloor+1.
\]
Similarly,
\[
|\Lambda_c|
=
\left\lfloor\frac{c}{2}\right\rfloor+1.
\]

Since $T_{r,c}=C_r\square  C_c$, every torus eigenvalue is the sum of one eigenvalue of $C_r$ and one eigenvalue of $C_c$. Therefore, the set of distinct torus eigenvalues is the sumset
\[
\Lambda_{T_{r,c}}
=
\Lambda_r+\Lambda_c
=
\{x+y:x\in\Lambda_r,\ y\in\Lambda_c\},
\]
and consequently the number of distinct torus eigenvalues is
\[
m(T_{r,c})
=
|\Lambda_r+\Lambda_c|.
\]
Importantly, this number need not equal $|\Lambda_r|\cdot|\Lambda_c|$, because different pairs $(x,y)$ can produce the same sum $x+y$. Thus, the number of distinct torus eigenvalues is controlled by the cardinality of the sumset rather than by the number of eigenvalue pairs.
For finite subsets $A,B\subset\mathbb{R}$,
\[
|A+B|\geq |A|+|B|-1.
\]

Using the expressions above for $|\Lambda_r|$ and $|\Lambda_c|$, we obtain
\[
m(T_{r,c})
\geq
\left\lfloor\frac{r}{2}\right\rfloor
+
\left\lfloor\frac{c}{2}\right\rfloor
+1.
\]
For the balanced tori considered here, $r$ and $c$ are even, so
\[
m(T_{r,c})
\geq
\frac{r+c}{2}+1.
\]

Applying the spectral-cardinality bound therefore gives
\[
\overline{\delta}_{\infty}(T_{r,c})
\leq
2\sin\left(
\frac{\pi}{m(T_{r,c})}
\right)
\leq
2\sin\left(
\frac{\pi}{(r+c)/2+1}
\right).
\]
Hence
\[
\overline{\delta}_{\infty}(T_{r,c})
=
O\left(\frac{1}{r+c}\right).
\]
Since $r,c=\Theta(\sqrt{N})$ for the balanced torus,
\[
\boxed{
\overline{\delta}_{\infty}(T_{r,c})
=
O\left(\frac{1}{\sqrt{N}}\right)
}.
\]

As in the Hamiltonian case, this provides only an upper bound. Establishing a corresponding lower bound for the long-time averaged unitary separation would require controlling the complete set of torus eigenvalue differences. Accordingly, the bounds above do not determine the exact asymptotic scaling of either torus metric, and we leave this as an open analytical question.

\subsection{Cycle versus Torus}
\label{sec:analytic-torus-vs-cycle}

For the cycle, the normalised Hamiltonian eigenvalue separation scales as
\[
\widetilde{\Delta}_{C_N}
=
\Theta(N^{-2}),
\]
while for the balanced 2-D torus we obtain only
\[
\widetilde{\Delta}_{T_{r,c}}
=
O(N^{-1}).
\]
Because the torus result is only an upper bound, it does not establish that
the torus has a larger minimum spacing than the cycle. The true torus spacing
may be smaller due to cancellations between contributions from its two cycle
spectra.

The same issue appears for the unitary metric:
\[
\overline{\delta}_{\infty}(C_N)
=
O(N^{-1}),
\qquad
\overline{\delta}_{\infty}(T_{r,c})
=
O(N^{-1/2}).
\]
These bounds therefore do not determine the asymptotic ordering of the cycle
and torus. Their relative performance at the finite system sizes considered
in this work is instead treated as an empirical observation.

\begin{table}[t]
    \centering
    \begin{tabular}{|l|l|l|}
    \hline
    \textbf{Topology}
    & \textbf{$\widetilde{\Delta}_{H_M}$}
    & \textbf{$\overline{\delta}_{\infty}(H_M)$} \\
    \hline

    Complete $K_N$
    & $\Theta(1)$
    & $\frac{4}{\pi}=\Theta(1)$ \\

    Hypercube $Q_q$
    & $\Theta(1/q)=\Theta(1/\log_2 N)$
    & $\Theta(1/q)=\Theta(1/\log_2 N)$ \\

    Cycle $C_N$
    & $\Theta(N^{-2})$
    & $O(N^{-1})$ \\

    2-D Torus $T_{r,c}$
    & $O(N^{-1})$
    & $O(N^{-1/2})$ \\

    \hline

    \end{tabular}

    \caption{Summary of the analytical spectral results. Exact asymptotic scaling is established for both metrics for the complete graph and hypercube, and for the Hamiltonian metric of the cycle. The listed cycle unitary result and both torus results are upper bounds only; corresponding matching lower bounds are not established.}
    \label{table:topology-comparison}
\end{table}

\section{Experimental Validation}
\label{sec:experiments}

\subsection{Setup}
\label{sec:experiments-setup}

We test whether the proposed spectral metrics are associated with QAOA
performance across four mixer topologies. To examine when this relationship
holds, we consider two qualitatively different classes of cost Hamiltonians:
a permutation-invariant random-energy ensemble and structured Ising QUBOs.

\paragraph{Unstructured instances.}
We use a permutation-invariant random-energy ensemble as a model of cost
landscapes with no preferred geometry in the computational basis. The energies
of different configurations are sampled independently, so neighbouring bit
strings are not expected to have similar costs and no mixer topology is
favored by alignment with the cost landscape.

This type of behavior is relevant, for example, to cryptographic search, where
the avalanche property decorrelates nearby inputs, and to the low-energy
spectrum of high-precision number partitioning, which is well described by a
Random Energy Model
\cite{mertens2000random,derrida1981random,borgs2001phase}.
The ensemble therefore provides a natural baseline for studying the influence
of the mixer topology itself.

For each $N\in\{16,32,64,128,256,512\}$ and $p\in\{1,\dots,4\}$, we generate $20$
instances with
\[
C(x)=c_x,
\qquad
c_x\sim\mathcal U(-1,1)
\]
independently for
$x\in\mathcal F=\{0,1\}^q$, with $|\mathcal F|=N=2^q$.

\paragraph{Ising QUBO instances.}
As a structured setting, we consider random 1-D and 2-D Ising QUBOs.
The Ising/QUBO form is widely used to represent combinatorial optimisation
problems~\cite{lucas2014ising}. Unlike the random-energy ensemble, these cost
functions contain local structure: flipping a single bit changes only the
terms involving that variable and its neighbours. The costs of nearby
configurations are therefore correlated, allowing us to test whether the
spectral ordering observed for unstructured costs persists when
mixer--cost alignment becomes relevant.

For
$q\in\{4,6,8,9\}$, corresponding to
$N=2^q\in\{16,64,256,512\}$, and
$p\in\{1,2,3,4\}$, we generate $20$ instances for each configuration.
For an interaction graph $G_C=(V,E)$ on the $q$ binary variables, the cost
function is
\[
C(x)
=
\sum_{(i,j)\in E} J_{ij}z_i z_j
+
\sum_i h_i z_i,
\]
where $z_i=1-2x_i$ and
$J_{ij},h_i\sim\mathcal U(-1,1)$ independently.

For the 1-D instances, nearest-neighbour variables are coupled on a ring with
periodic boundary conditions,
\[
E=\{(i,i+1\bmod q):i=0,\ldots,q-1\}.
\]
For the 2-D instances, nearest-neighbour variables are coupled on an
$r\times c$ lattice with $rc=q$ and periodic boundary conditions in both
directions.

The Ising interaction graph and the mixer graph play different roles.
The interaction graph contains $q$ vertices and defines the cost Hamiltonian,
whereas the mixer topologies---cycle, torus, complete graph, and hypercube---are
defined over the full $N=2^q$ computational-basis states.

For structured costs, the assignment of computational-basis states to mixer
vertices can affect the QAOA dynamics, particularly for the cycle and torus.
The indexing convention used in our experiments is therefore provided with
the implementation details in the accompanying code (see Code Availability).

\paragraph{Optimization.}
Because the mixer families have different parameter-domain structures, we use
different search strategies while keeping the objective-evaluation budget
fixed at approximately $2000$ evaluations per run:
\begin{itemize}[leftmargin=*, itemsep=2pt]
    \item \textbf{Cycle and torus:}
    multi-start Nelder--Mead with Cauchy-distributed restarts and local
    refinement.
    \item \textbf{Complete graph and hypercube:}
    uniform random search over $[0,\pi]^{2p}$.
\end{itemize}

Since different optimisation procedures are used across mixer families,
cross-family comparisons may contain optimiser-dependent effects despite the
matched evaluation budget. The complete graph and hypercube, however, use the
same optimisation procedure and can therefore be compared without this
particular source of variation.

\paragraph{Performance metric.}
We measure QAOA performance relative to the uniform-state baseline and the
exact optimum. Let
\[
C_{\mathrm{avg}}
=
\frac{1}{|\mathcal{F}|}
\sum_{x\in\mathcal{F}} C(x)
\]
be the average cost over all basis states, which is also the expected cost of
the initial uniform state $\ket{\psi_0}$. We define the approximation score
\[
r =
\frac{C_{\mathrm{avg}}-M_p}
     {C_{\mathrm{avg}}-C_{\min}},
\]
where $M_p$ is the best QAOA expectation value obtained at depth $p$, and
\[
C_{\min}=\min_x C(x)
\]
is the exact optimum.

With the uniform-state baseline included, $r=0$ corresponds to the initial
uniform state and $r=1$ to the exact optimum. We report the mean and standard
deviation of $r$ over the $20$ instances for each
$(N,p,\text{topology})$ configuration and cost family.

\subsection{Results: Unstructured Costs}
\label{sec:experiments-results-unstructured}

\begin{figure}[!htbp]
    \centering
    \includegraphics[width=\textwidth]{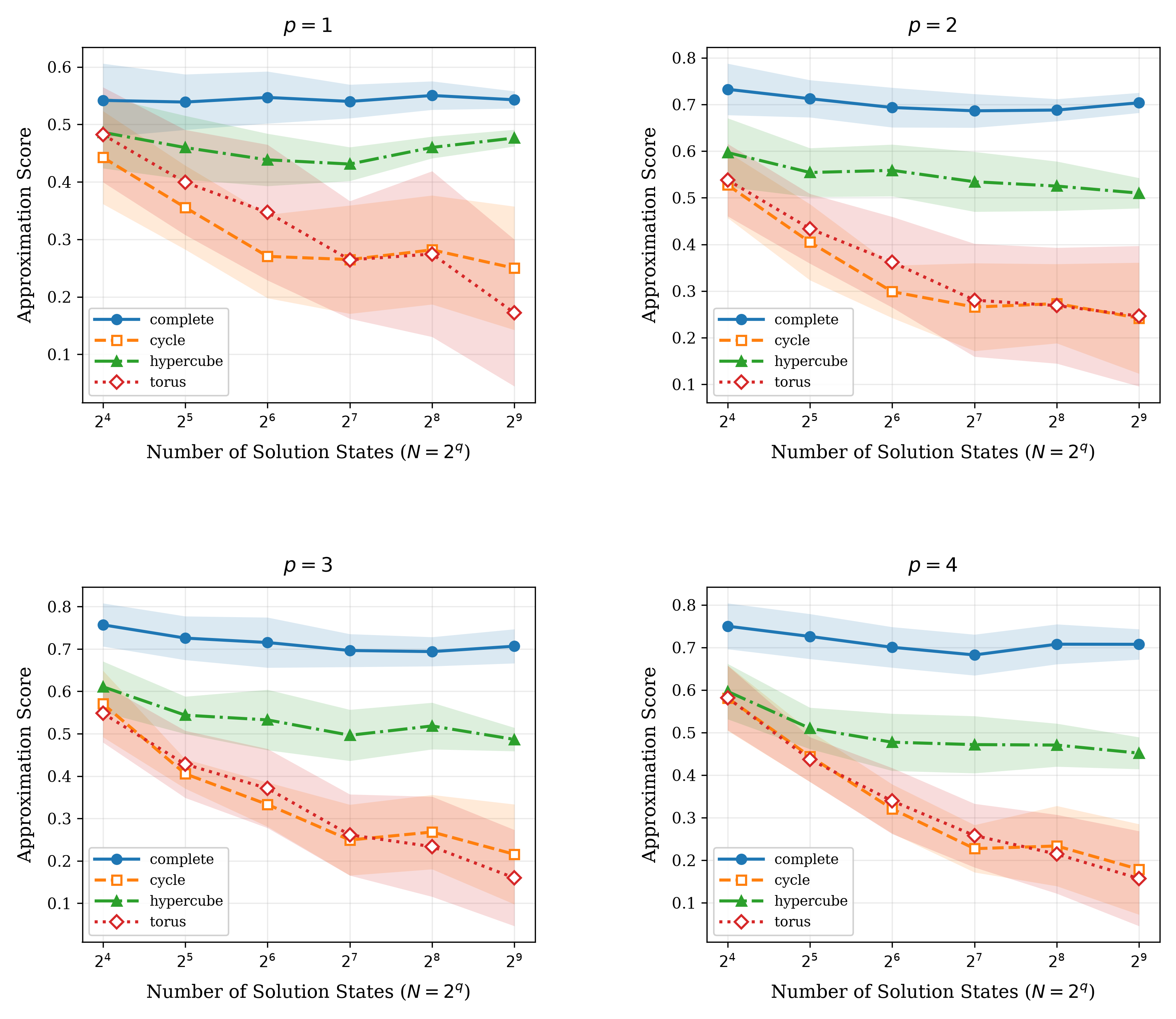}
    \caption{Mean approximation score $r$ as a function of the solution-space size
    $N=2^q$, shown separately for QAOA depths $p=1,\ldots,4$, on the
    permutation-invariant random-energy ensemble. Shaded regions indicate
    one standard deviation over the $20$ instances.}
    \label{fig:random-vs-N}
\end{figure}

\begin{figure}[!htbp]
    \centering
    \includegraphics[width=\textwidth]{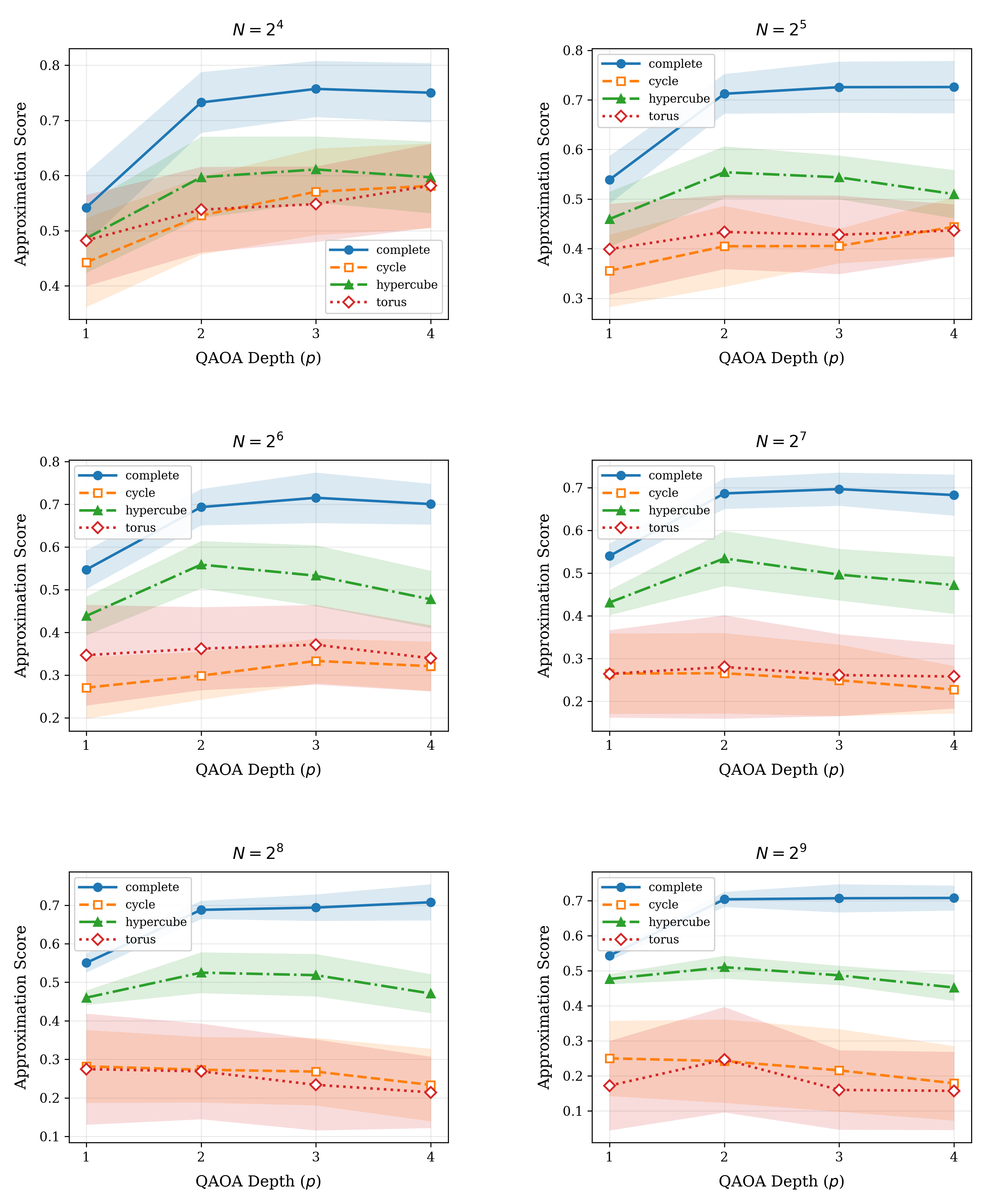}
    \caption{Mean approximation score $r$ as a function of QAOA depth $p$,
    shown separately for each solution-space size $N=2^q$, on the
    permutation-invariant random-energy ensemble. Shaded regions indicate
    one standard deviation over the $20$ instances.}
    \label{fig:random-vs-p}
\end{figure}

Figures~\ref{fig:random-vs-N} and~\ref{fig:random-vs-p} show the QAOA performance on the random-energy ensemble. Across the tested system sizes and depths, the complete-graph mixer achieves the highest mean approximation score, while the hypercube is generally the second-best performing family. The cycle and torus mixers show a much stronger decline in performance as the size of the solution space increases.

This behaviour is broadly consistent with the spectral hypothesis developed in the preceding sections. The complete graph has $\widetilde{\Delta}_{K_N}=\Theta(1)$ and $\overline{\delta}_{\infty}(K_N)=\Theta(1)$, while the hypercube shows a slower decay of $\Theta(1/\log N)$ in both spectral quantities. In contrast, the cycle has $\widetilde{\Delta}_{C_N}=\Theta(N^{-2})$ and $\overline{\delta}_{\infty}(C_N)=O(N^{-1})$, while the balanced torus satisfies the upper bounds $\widetilde{\Delta}_{T_{r,c}}=O(N^{-1})$ and $\overline{\delta}_{\infty}(T_{r,c})=O(N^{-1/2})$. The empirical separation between the complete/hypercube mixers and the cycle/torus mixers is therefore consistent with the broad spectral separation between these topology classes.

The analytical results do not, however, establish an asymptotic ordering between the cycle and torus. The unitary results for both are only upper bounds, and the torus Hamiltonian result is also an upper bound. Their relative performance in these figures should therefore be viewed as a finite-size empirical observation rather than as a consequence of the asymptotic analysis.

The same broad ordering is observed across all tested circuit depths $p\in\{1,2,3,4\}$. Increasing the depth generally gives the largest improvement from $p=1$ to $p=2$, after which the performance often saturates or decreases slightly. Since the optimisation budget is fixed while the number of variational parameters increases with $p$, this should not be interpreted as evidence that greater QAOA depth intrinsically reduces solution quality. It may also reflect the greater difficulty of parameter optimization at larger $p$.

Overall, the random-energy experiments support the hypothesis that, when the cost Hamiltonian does not impose a preferred configuration-space geometry, the proposed topology-level spectral quantities can provide useful qualitative information about mixer performance.

\subsection{Results: Structured Ising QUBOs}
\label{sec:experiments-results-ising}

\begin{figure}[htbp]
    \centering
    \includegraphics[width=0.8\textwidth]{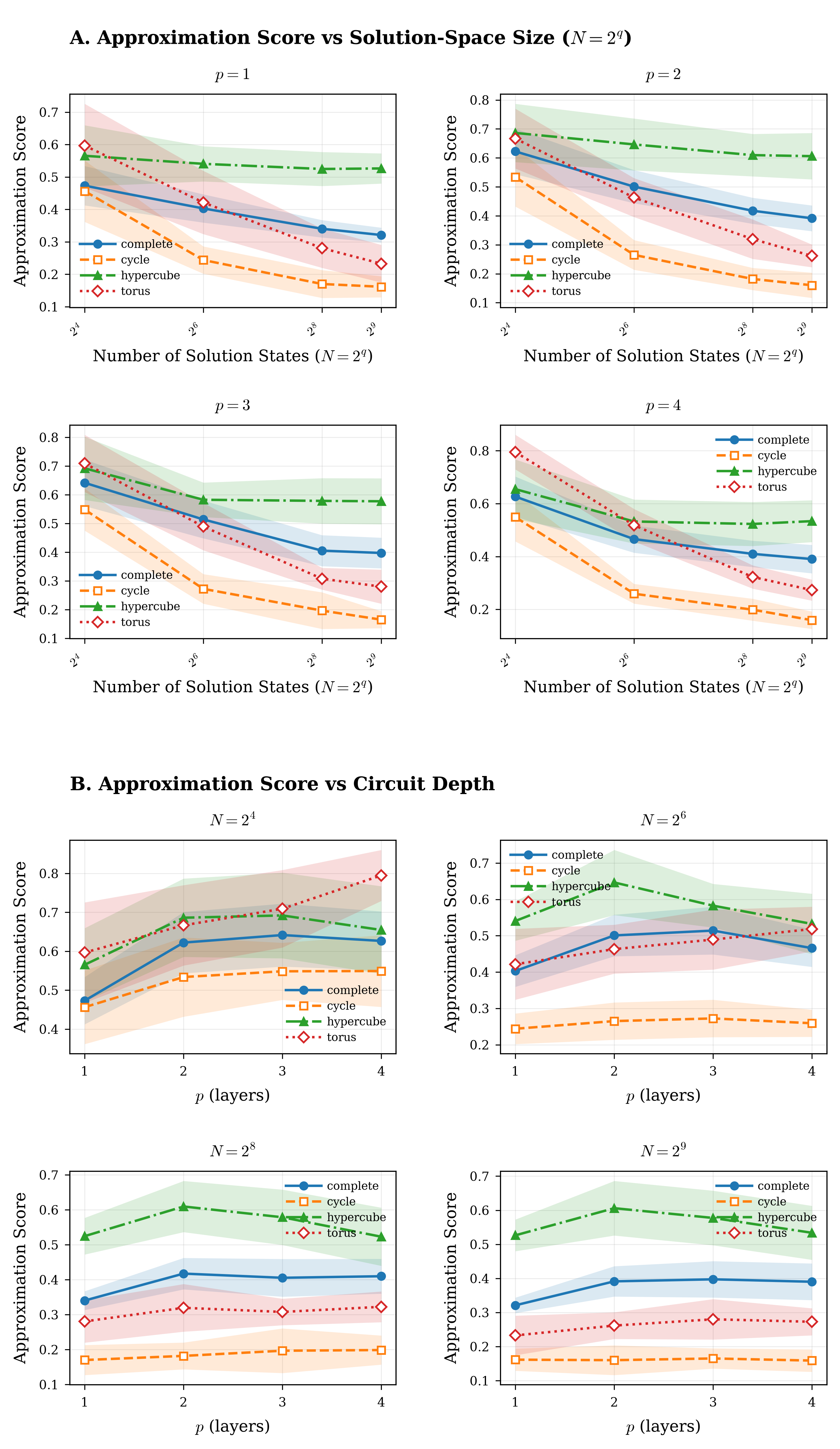}
    \caption{QAOA performance on the structured 1-D Ising instances. The upper panels show the mean approximation score as a function of solution-space size $N=2^q$ for each depth $p$, while the lower panels show its dependence on $p$ for fixed $N$. Shaded regions indicate one standard deviation over the $20$ instances.}
    \label{fig:ising-1d}
\end{figure}

\begin{figure}[htbp]
    \centering
    \includegraphics[width=0.8\textwidth]{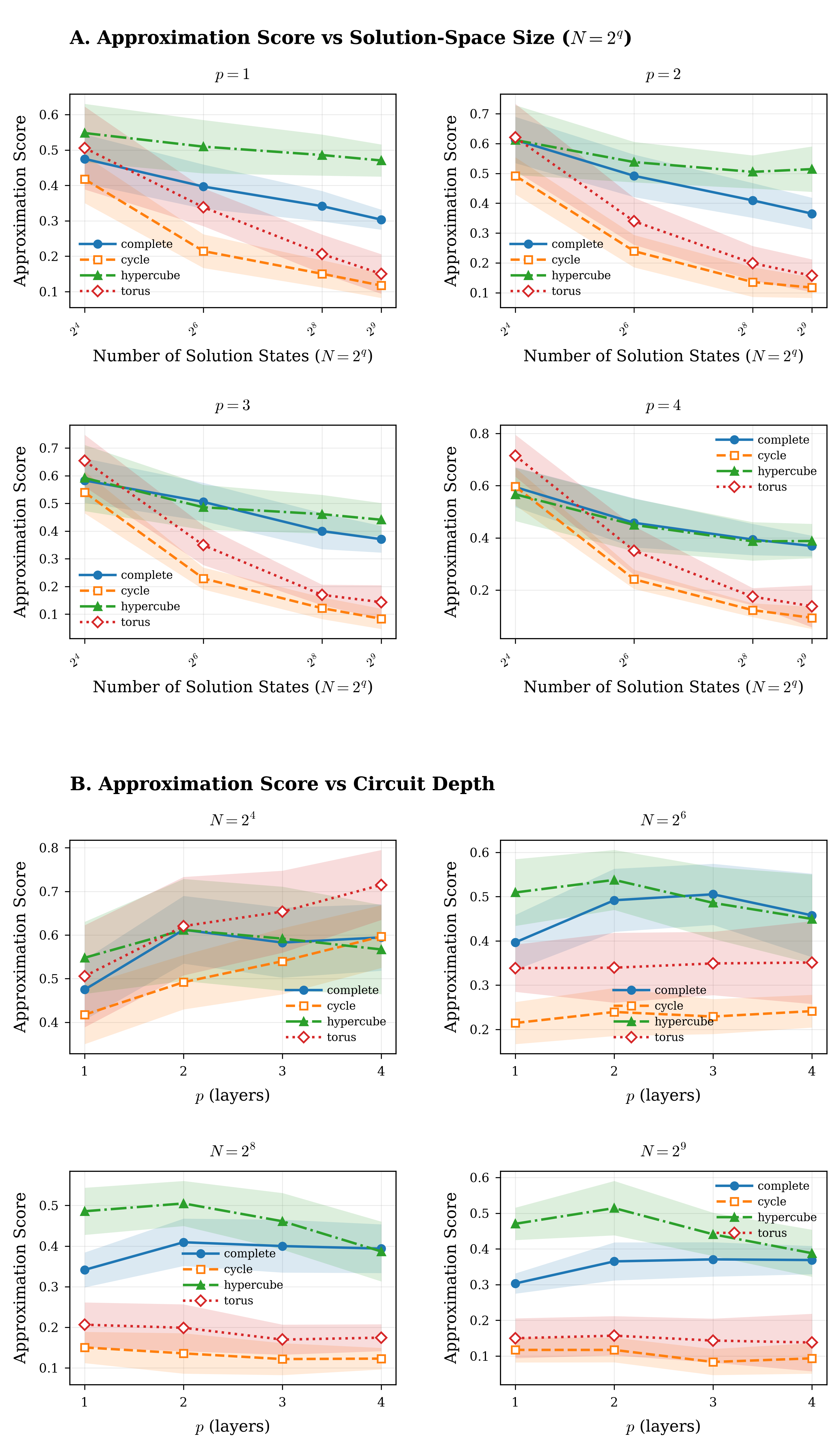}
    \caption{QAOA performance on the structured 2-D Ising instances. The upper panels show the mean approximation score as a function of solution-space size $N=2^q$ for each depth $p$, while the lower panels show its dependence on $p$ for fixed $N$. Shaded regions indicate one standard deviation over the $20$ instances.}
    \label{fig:ising-2d}
\end{figure}

Figures~\ref{fig:ising-1d} and~\ref{fig:ising-2d} show the corresponding results for the structured 1-D and 2-D Ising cost families. Unlike the random-energy ensemble, the ordering suggested by the cost-agnostic spectral metrics does not persist. As the system size increases, the hypercube generally matches or outperforms the complete-graph mixer in both the 1-D and 2-D experiments, even though the complete graph has larger values of both spectral metrics.

This difference becomes clearer for the larger solution spaces. The performance of the complete graph decreases with increasing $q$, while the hypercube retains substantially higher approximation scores across the tested depths. The cycle performs poorly at larger sizes, while the torus generally lies between the cycle and the complete or hypercube mixers, depending on the instance family and circuit depth. At the smallest sizes, the differences are less systematic, and several error bands overlap, so this ordering should not be interpreted as universal for every $(q,p)$ configuration.

A plausible explanation for the strong hypercube performance is its local connectivity in configuration space. Hypercube edges connect bit strings that differ in exactly one bit. For a local Ising cost Hamiltonian, a single-bit flip changes only the local-field term for that spin and the interaction terms involving it. The corresponding mixer moves therefore make local changes to the structured cost landscape. In contrast, the complete graph connects every pair of computational-basis configurations, including states that differ by many simultaneous bit flips. These results suggest that mixer--cost alignment can become important for structured problems, and that spectral separation alone is not sufficient to determine QAOA performance.

The $q=4$ case provides a useful example of this limitation. Here, the balanced torus is $T_{4,4}=C_4\square C_4$, which is isomorphic as an abstract graph to the four-dimensional hypercube $Q_4$. The two mixers therefore have identical adjacency spectra and receive identical values under both spectral metrics. Nevertheless, their QAOA performance on the structured Ising instances can differ noticeably. This does not contradict the graph isomorphism: the isomorphism corresponds to a permutation of the computational-basis vertices, while a fixed structured cost Hamiltonian need not be invariant under that permutation. This example illustrates that eigenvalue-based metrics alone do not capture the alignment between mixer connectivity, basis-state labelling, and the cost landscape.

Taken together, these experiments support the intended scope of the proposed metrics. For the permutation-invariant random-energy ensemble, where the computational-basis labels carry no preferred structure, the empirical performance is broadly consistent with the topology-level spectral characterisation. For structured Ising costs, this correspondence weakens or reverses, indicating that additional information about the relationship between the mixer and the cost Hamiltonian is needed.

\section{Discussion and Future Work}
\label{sec:discussion}

Our results reveal two distinct regimes for the spectral heuristics
$\widetilde{\Delta}_{H_M}$ and
$\overline{\delta}_{\infty}(H_M)$.
For the permutation-invariant random-energy ensemble, the empirical ordering is
\[
K_N > Q_q > T_{r,c} \approx C_N.
\]
This is broadly consistent with the analytical spectral characterisation in
Sections~\ref{sec:analytic-complete}--\ref{sec:analytic-torus}: the complete
graph and hypercube lie in a more strongly separated spectral regime than the
cycle and torus. The available bounds do not fully determine the asymptotic
ordering between the cycle and torus, so their relative ordering should be
viewed as a finite-size empirical observation. Overall, these results support
our working hypothesis that, when the cost Hamiltonian imposes no preferred
geometry on the computational basis, topology-level spectral properties can
provide useful qualitative information about QAOA mixer performance.

For structured Ising QUBOs, this ordering breaks down. The hypercube mixer
generally matches or outperforms the complete graph despite having smaller
spectral separations. This illustrates the limitation discussed in
\cref{sec:intro-scope}: a cost-agnostic spectral characterisation need not
remain predictive once the cost Hamiltonian introduces exploitable structure.
Hypercube edges connect computational-basis states that differ by a single bit
flip. For a local Ising Hamiltonian, such a flip modifies only the local-field
term of that spin and the interaction terms involving it. The resulting local
moves can therefore be better aligned with the Ising cost landscape than the
all-to-all connectivity of the complete graph.

More generally, the proposed metrics depend only on the eigenvalues of the
mixer Hamiltonian and therefore cannot capture all aspects of mixer--cost
compatibility. For example, if two mixer Hamiltonians are related by a
permutation of the computational basis,
\[
H'_M=P H_M P^\dagger,
\]
then they have identical spectra and consequently identical values of both
$\widetilde{\Delta}_{H_M}$ and
$\overline{\delta}_{\infty}(H_M)$. Nevertheless, for a fixed structured cost
Hamiltonian $H_C$, their QAOA dynamics can differ because the permutation
changes which computational-basis configurations are connected by the mixer.
Thus, eigenvalue information alone cannot fully characterise the alignment
between mixer connectivity and the cost landscape.

These observations motivate richer mixer-selection frameworks that combine
intrinsic spectral properties of the mixer with information about the
structure of the cost Hamiltonian. Related approaches that incorporate problem
structure into mixer design have begun to be explored
~\cite{matwiejew2024quantum}. Our empirical validation is also limited to
random-energy and Ising-type Hamiltonians, and it remains to be seen how well
the proposed heuristics generalise to other optimisation problems and cost
structures.

Several directions for future work follow from these results.
First, the exact asymptotic minimum spectral separation of the 2-D torus
remains unresolved. For the balanced torus, we obtain
\[
\widetilde{\Delta}_{T_{r,c}}=O(N^{-1}),
\qquad
\overline{\delta}_{\infty}(T_{r,c})=O(N^{-1/2}),
\]
but neither result currently has a matching asymptotic lower bound. Since
torus eigenvalue differences involve sums of two cycle-spectrum differences,
near-cancellations may lead to smaller spacings. Establishing matching lower
bounds would determine the exact asymptotic scaling of the torus metrics and
clarify their relation to those of the cycle.

Second, the framework can be extended to constrained mixers and other graph
families, particularly cases where
$\widetilde{\Delta}_{H_M}$ and
$\overline{\delta}_{\infty}(H_M)$ exhibit qualitatively different behavior.
Such studies may help identify which aspects of spectral structure are most
relevant to QAOA performance.

Third, and most importantly, the breakdown of the cost-agnostic spectral
ordering on structured problems motivates a unified framework that incorporates
both spectral characteristics and mixer--cost alignment. Such a framework may
need information beyond mixer eigenvalues, including eigenvectors,
computational-basis connectivity, or quantities that directly compare the mixer
and cost Hamiltonians.

Further empirical work should test a broader range of cost ensembles and
optimisation problems, and examine the robustness of the observed topology
ordering under common parameter-optimisation strategies. Establishing a formal
connection between the proposed spectral metrics and shallow-depth QAOA
approximation ratios also remains an important open theoretical question.

\section{Conclusion}
\label{sec:conclusion}

We introduce two complementary spectral metrics for QAOA mixers: the
operator-norm-normalized minimum Hamiltonian eigenvalue separation
$\widetilde{\Delta}_{H_M}$ and the long-time averaged minimum unitary
eigenvalue separation $\overline{\delta}_{\infty}(H_M)$. Motivated by
quantum-walk mixing theory, these provide scale-independent Hamiltonian- and
unitary-level characterisations of mixer spectral structure, which we use as
heuristic indicators of QAOA mixer performance.

We derive exact expressions and asymptotic bounds for the complete graph,
hypercube, cycle, and balanced 2-D torus. These results reveal a broad spectral
separation between the mixer families, although matching asymptotic bounds
remain unresolved for the torus metrics and for the long-time unitary metric
of the cycle. On the permutation-invariant random-energy ensemble, the
empirical QAOA performance is broadly consistent with this spectral
characterisation across the tested system sizes and circuit depths. This
supports the use of topology-level spectral quantities as qualitative
mixer-selection heuristics when the cost Hamiltonian does not impose a
preferred geometry on the configuration space.

The structured 1-D and 2-D Ising QUBO experiments show the limitations of
this cost-agnostic characterisation. The hypercube generally matches or
outperforms the complete graph despite having smaller spectral separations.
This indicates that mixer--cost alignment can become important when the cost
Hamiltonian contains exploitable structure, and that mixer eigenvalues alone
are not sufficient to characterise QAOA performance in such settings.
Developing structure-aware criteria that combine intrinsic mixer spectral
properties with information about the cost landscape therefore remains an
important direction for future work.

\paragraph{Code Availability.}

The code used to generate the problem instances, implement the QAOA
simulations and mixer topologies, perform parameter optimisation, and produce
the numerical results reported in this work is available from the authors
upon reasonable request.

\paragraph{Declaration of AI-Assisted Writing.}

The scientific ideas, methodology, mathematical analysis, code development
and implementation, experimental design, and interpretation of the results
presented in this work were carried out by the authors. Generative AI tools
were used only to assist with parts of the manuscript writing process,
including improving language, grammar, clarity, and presentation. The authors
reviewed the final manuscript and take full responsibility for its content.

\bibliographystyle{plainurl}
\bibliography{references}

\end{document}